\documentclass{vldb}
\usepackage{graphicx}
\usepackage{algorithm}
\usepackage[noend]{algpseudocode}
\usepackage{balance}  
\usepackage{xspace}
\usepackage{booktabs}
\usepackage{cleveref}

\newcommand{\sys}{\textsf{DQM}\xspace}
\newtheorem{problem}{Problem}

\makeatletter
\def\@copyrightspace{\relax}
\makeatother

\usepackage{tcolorbox}
\usepackage{pgfkeys}

\definecolor{pop}{RGB}{254,81,0}

\usepackage{listings}
\usepackage{url}

\newcommand{\rulebox}[5]{
\begin{tcolorbox}[capture=minipage,left=0pt,right=1pt,top=1pt,bottom=1pt,boxrule=0.5pt,coltitle=black]
\setlength{\tabcolsep}{2pt}
\begin{tabular}{ ll }
\textsf{GIVEN}: & {#3}\\
\textsf{RETURNS}: & {#4}\\
\end{tabular}
\end{tcolorbox}
}

\newcommand{\mdpbox}[5]{
\begin{tcolorbox}[capture=minipage,left=0pt,right=1pt,top=1pt,bottom=1pt,boxrule=0.5pt,coltitle=black]
\setlength{\tabcolsep}{2pt}
\begin{tabular}{ ll }
\textsf{State}: & {#1}\\
\textsf{Action}: & {#2}\\
\textsf{Reward}: & {#3}\\
\textsf{Policy}: & {#4}\\
\end{tabular}
\end{tcolorbox}
}

\begin{document}


\title{Opportunistic View Materialization with Deep Reinforcement Learning
}

\numberofauthors{3} 

\author{
%
%
\alignauthor
Xi Liang \\
       \affaddr{University of Chicago}\\
       \email{xiliang@uchicago.edu}
\alignauthor
Aaron J. Elmore \\
       \affaddr{University of Chicago}\\
       \email{aelmore@uchicago.edu}
\alignauthor
Sanjay Krishnan \\
       \affaddr{University of Chicago}\\
       \email{skr@uchicago.edu}
}

\maketitle

\begin{abstract}
Carefully selected materialized views can greatly improve the performance of OLAP workloads.
We study using deep reinforcement learning to learn adaptive view materialization and eviction policies.
Our insight is that such selection policies can be effectively trained with an asynchronous RL algorithm, that runs paired counter-factual experiments during system idle times to evaluate the incremental value of persisting certain views.
Such a strategy obviates the need for accurate cardinality estimation or hand-designed scoring heuristics.
We focus on inner-join views and modeling effects in a main-memory, OLAP system.
Our research prototype system, called \sys, is implemented in SparkSQL and we experiment on several workloads including the Join Order Benchmark and the TPC-DS workload.
Results suggest that: (1) \sys can outperform heuristic when their assumptions are not satisfied by the workload or there are temporal effects like period maintenance, (2) even with the cost of learning, \sys is more adaptive to changes in the workload, and (3) \sys is broadly applicable to different workloads and skews.

\end{abstract}

\section{Introduction}
Carefully selected materialized views can greatly improve the performance of OLAP workloads.
We explore opportunistic materialization (OM)~\cite{lefevre2014opportunistic}, where a database preemptively caches important query (or sub-query) results for future use.
In an ideal world, OLAP systems would aggressively persist any result that could possibly be useful in the future. 
However, practical systems have resource constraints and usage patterns that are constantly evolving, and results that seem currently relevant can fall into disuse in the future.
Furthermore, maintaining a very large number of views can place a burden on query optimizer to select which views to use for a given query.
Therefore, the core technical problem in the design of OM systems is straight-forward: an effective dynamic view creation and eviction policy under storage constraints.

The database community has extensively studied view recommendation systems that take in a historical query workload, a database schema, and possibly a cost model to recommend the best views to create~\cite{perez2014history, dageville2004automatic, bruno2005automatic, papadomanolakis2007efficient, agrawal2000automated, zilio2004recommending}.
Such techniques are \emph{retrospective} because one implicitly assumes that future queries and data are similar to what was seen in the past, and choices that are good in retrospect are likely to be good in the future.
While retrospective approaches are principled, in the sense that they optimize a well-defined criterion, they suffer the obvious limitations~\cite{kotidis2001case}: they will not perform well for ad hoc queries, evolving workloads, or storage constraint changes. 

Truly dynamic strategies adapt to changing environments through eviction and re-creation actions. Existing dynamic view selection work is largely heuristic based. They define scoring criteria to quantify the value of keeping a view materialized ranging from the simplest Least-Recently-Used or Least-Frequently-Used approaches~\cite{deshpande1998caching, kotidis1999dynamat, scheuermann1996watchman} to  more sophisticated cost-model based approaches~\cite{perez2014history,nagel2010recycling}.
However, with LRU and the known ``sequential flooding'' failure case, existing dynamic policies are brittle by nature and can have subtle blind spots. 

\begin{figure}[t]
    \centering
    \includegraphics[scale=0.275]{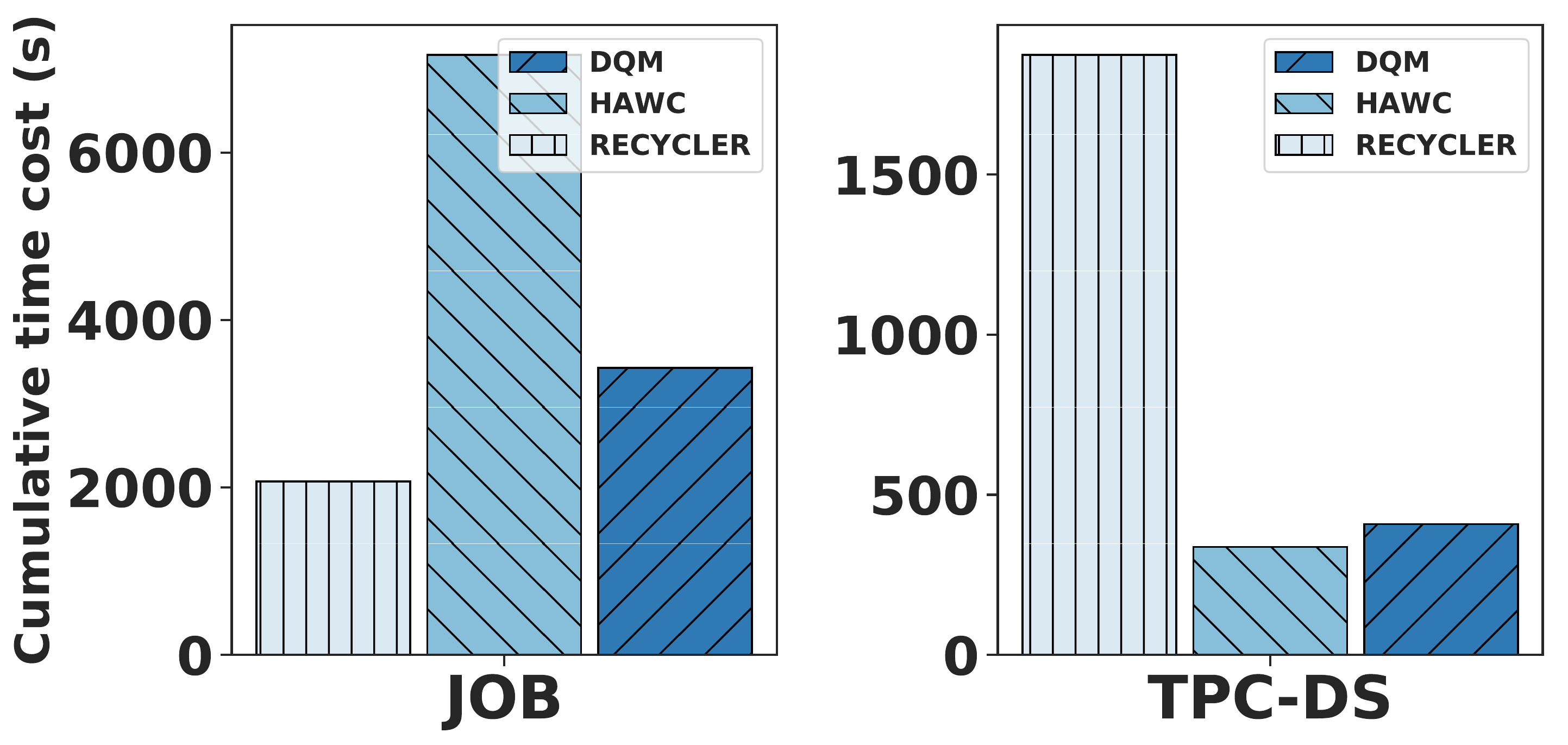}
    \caption{We compare state-of-the-art dynamic view selection criteria, HAWC and Recycler, on SparkSQL for the Join Order Benchmark and TPC-DS based workloads and the same storage constraint. HAWC and Recycler are sensitive to their particular heuristics and behave very differently on the two workloads. Our approach, \sys, learns a predictive model based on actual runtime feedback so it is more robust and outperforms both heuristics on overall time cost of the two workloads. \sys is competitive with these baselines even when learning time is included in the cumulative cost.}
    \label{paraintro}
\end{figure}

Figure \ref{paraintro} compares two state-of-the-art dynamic criteria, HAWC~\cite{perez2014history} and Recycler~\cite{nagel2010recycling}, on two different workloads of 1000 queries (derived from the Join Order Benchmark and TPC-DS).  
We implemented both heuristics in SparkSQL.
HAWC prioritizes usage frequency and Recycler prioritizes costly views (more details of the baselines and workloads we use can be found in Table \ref{workloads} and Table~\ref{baselines} in Section~\ref{sec:workloads}). We observe drastic performance differences across the benchmarks. 
Recycler works well when it can very significantly improve a small number of expensive queries, as in the Join Order Benchmark where a few queries involve $>10$ way joins. 
Exactly the opposite happens on TPC-DS, where it persists superfluous, large views and actually hurts performance.

Feedback from real execution times can ameliorate this brittleness.
Rather than relying on a heuristic, we can observe whether the creation of a view has a net positive or negative impact on subsequent query latencies.
This is fundamentally a learning problem as beneficial decisions should be remembered for the future and adverse decisions should be avoided.
This broad idea is inspired by recent works in query optimization, where actual runtimes are used to inform/optimize future plans~\cite{markl2003leo, chaudhuri2008pay, marcus2018deep, krishnan2018learning}.
To achieve this goal, we need a learning framework that can automatically and continuously update its model based on delayed observations.

Reinforcement Learning (RL) studies techniques that learn how to control general stateful systems (e.g., a database with persisted views)~\cite{sutton2018reinforcement, sutton1992reinforcement}.
A working definition of RL is ``learning by doing''; the algorithm takes actions and observes feedback via a performance metric (e.g., query runtime).
It assigns credit or blame to actions based on the feedback, and can even account for delayed effects.
As more feedback is observed, the learned behavior is increasingly informed.
In principle, an RL approach can learn which new views are valuable to materialize given the current system state (what views are currently persisted). 
This learning process can automatically adapt to changing workloads based on the observed performance.
It does not require a hard-coded heuristic nor does it need to explicitly generate an anticipated workload---it's predictive model is simply a ``means-to-an-end'' in terms of minimizing overall query latency.
The caveat is that has to be able to assign credit or blame to good and bad decisions it makes purely from how the queries execute.

This is the crux of the algorithmic challenge in applying RL in OM systems---ascertaining the net benefit of a view is difficult. 
Any system either makes a choice to use a view or not during query optimization, and the learning agent only observes the final runtime of one of these choices--and does not know the marginal effect with respect to the other choice.
We lack the ``paired'' experiment, where we observe the same query with and without the view, thereby quantifying the \emph{reward} of creating a view~\cite{box1978statistics}.
If the same queries (or similar queries) do not frequently repeat, the amount of time needed to learn an effective and adaptive materialization policy will be prohibitive.

Our insight is that OM systems need a new type of asynchronous RL algorithm that runs such paired experiments in the background. 
For every query, the system identifies a set of eligible views that can be opportunistically created. 
The scope of the current work is to focus on inner join views, but the technique is more general.
The system proactively takes the decision it thinks is best at the time using its query optimizer (possibly using no views). The counter-factual decision(s), the ones that the system did not take, are queued into an experiment buffer.
We simplify the experimentation problem by assuming an in-memory database with no extraneous unobserved state (e.g., the buffer pool state or caching effects).
Therefore, we can independently schedule and run these experiments during idle times producing retroactive marginal utility metrics for each view.
Our system can further model view refresh costs, but is not optimized for OLTP systems where these refresh events might be very frequent.

We implement this model in a prototype OM system called Deep Q-Materialization (\sys). \sys contains three main components: (1) An online query miner that analyzes a trace of SQL queries to identify candidate views for the current query to opportunistically materialize, (2) a RL agent that selects from the set of candidates, and (3) an eviction policy that selects views to delete. \sys is integrated with SparkSQL. The adaptive policy interacts with the Spark environment through a RESTful API and can easily be ported to other SQL-based data processing systems. 

Figure \ref{paraintro} shows the potential advantage of \sys.
Over workloads of 1000 queries \sys is competitive with the best of  the heuristics on each workload in terms of cumulative query latency. This is even including the time needed to learn the selection model.
Further experiments find that \sys can match or outperform standard heuristics policies across 5 different temporal query patterns on two different workloads. 
\sys maximizes utilization of available storage for the given workload and query processing engine.  

In summary, this paper makes the following contributions:
\begin{itemize}
    \item We formalize online view selection in opportunistic materialization systems as a Markov Decision Process (MDP).
    
    \item We propose a new asynchronous reinforcement learning algorithm, based on the Double DQN model, to optimize this MDP objective online.

    \item We propose a new credit-based eviction model that can enforce a hard storage constraint on views created by the learned selection policy.
    
    \item We compare our approach to classical and state-of-the-art baselines to demonstrate \sys's adaptivity, latency, and robustness.
\end{itemize}

The remainder of this paper is structured as follows. Section 2 gives an overview of related work. Section 3 presents our problem setting and system architecture. In section 4, we discuss technical details of our reinforcement learning approach. Section 5 describes the design of our eviction policy. Section 6 presents the experimental evaluation of our system. Finally, Section 7 concludes the paper with discussion.

\section{Background}
Here we overview prior approaches to OM, discuss relevant reactive and predictive policies, and review reinforcement learning.

\subsection{Motivation and Applications}
We borrow the term ``opportunistic materialization''~\cite{lefevre2014opportunistic} that describes automatic persistence in large-scale data processing systems like Hive and Pig. While very different from our work (the aforementioned persistence was for intra-task optimization), we use the term \emph{opportunistic} to describe any materialization that is an artifact of execution and not explicitly defined by a human database administrator. 

Reusing previously computed intermediate results across queries can significantly improve overall throughput and latency.
While the general idea has been studied before, we believe there are several trends that encourage us to revisit this problem.
First, due to advances in natural language processing and computer vision, compute-bound UDFs for machine learning inference are increasingly common.
Avoiding additional re-computation of previously computed values can greatly improve query processing.
Second, the growth in cloud-based database offerings provide individual users with larger storage constraints and more flexibility to materialize a large number of views.
Finally, batch data processing systems like SparkSQL are increasingly fast enough for ad hoc query processing and are used in a data exploration context where many related queries are executed in quick succession.

Despite this new need, existing heuristic-based approaches are limited.  As the example in Figure \ref{paraintro} shows, there is no ``one-size fits all" heuristic. 
It is challenging to decide \emph{a priori} whether a heuristic-based approach will even work for a workload.
More subtly, these heuristics can be at the mercy of the DBMS's cost estimation and query optimizer and actually hurt performance.
 We believe this is an opportunity for adaptive online approaches like \sys that use actual observed query latencies as feedback.

\subsection{Retrospective Policies}
The classical approaches of static view selection, which select the best materialized views from a given input candidate view set under storage and/or maintenance constraints, are one extreme of the design space for OM.
These approaches recommend the best views to create based on a query workload and storage constraints~\cite{perez2014history, dageville2004automatic, bruno2005automatic, papadomanolakis2007efficient, agrawal2000automated, zilio2004recommending}.

In these approaches, ``optimality'' is well-defined.
They find the best set of views that satisfy the storage constraint and that most improve the estimated query execution cost of the entire workload.
The downside is that one only searches over ``static'' strategies, where the views are created upfront.
These systems have difficulty reasoning about evicting and re-creating views.
Even if we were to periodically run such view recommendation tools, we would have a number of difficult, unresolved questions: how to window the workload, how to penalize view creation costs, and how frequently to re-run a recommendation tool.

\subsection{Reactive and Predictive Policies}
Dynamic strategies ostensibly address these issues.
DynaMat~\cite{kotidis1999dynamat} and WATCHMAN~\cite{scheuermann1996watchman} were seminal projects in dynamic materialized view management. We term these approaches ``reactive'' because rather than purely relying on a historical workload, they react to transient usage patterns. Systems in this space have to solve multiple problems: what views to materialize~\cite{mami2012survey,phan2008}, when to evict views~\cite{deshpande1998caching}, and how to select which views to use~\cite{chen1994implementation}.  Older systems borrowed strategies from database paging (e.g., LRU), and state-of-the-art systems apply more sophisticated scoring heuristics that account for creation and usage costs.
HAWC~\cite{perez2014history} scores views based on a cost-model and maintains a table of such scores for persisted views. New views that have a higher score than those in the table force an eviction event. The scores in the table are windowed to consider only the latest $K$ queries to ensure adaptivity. RECYCLER~\cite{recycler6544837} prioritizes the most expensive views (in terms of creation cost). The reasoning being that these views are harder to re-create if they are evicted.
An intriguing variant of these ideas is to form a predictive model that forecasts the type and distribution of queries one may encounter~\cite{ma2018query}.
As far as we can tell, such a predictive approach has not yet been truly applied to materialized view selection (Ma et al. only study index creation) and defer a detailed exploration of workload forecasting for future work.

\subsection{Reinforcement Learning}
Reinforcement Learning (RL) studies algorithms that learn how to control a stateful system.
In RL, a hypothetical learning agent takes decisions to affect the state of the system.
After each decision the system updates its state (possibly non-deterministically), and then, the agent observes a ``reward'', or a score of how good that decision is.
The objective for the agent is to learn a \emph{policy}, a function that automatically takes a decision based on the current state, with the maximum long term reward.

Mathematically, the interaction between the agent and the system is described by a Markov Decision Process (MDP), which is a 6-tuple $\langle \mathcal{S}, \mathcal{A}, p_0, p, R, \gamma \rangle$, where $\mathcal{S}$ denotes the state space (the set of all possible states), $\mathcal{A}$ the action space (set of all possible decisions), $p_0$ the initial state distribution (how the system starts out), $p(s_{t+1} \mid s_{t}, a_{t})$ the state transition distribution (how the state changes given a decision), $R(s_t, a_t) \in \mathbb{R}$ is the reward function, and $\gamma \in [0,1)$ the discount factor (a weight to discount future rewards). 
The objective of an MDP is to find a decision policy, a probability distribution over actions $\pi: \mathcal{S} \mapsto \Delta(A)$.
A policy $\pi$ induces the distribution over trajectories $\xi = [(s_0,a_0),(s_1,a_1),...,(s_N,a_N)]$:
\[
P_\pi(\xi) = p_0(x_0) \prod_{t=0}^{T-1} \pi(a_t \mid s_t) p(s_{t+1} \mid s_{t}, a_{t}).
\]
The \emph{value} of a policy is its expected total discounted reward over trajectories
\[
V_\pi = \mathbf{E}_{\xi \sim P_\pi}\left[\sum_{t=0}^{T-1} \gamma^t R(s_t,a_t)\right].
\]
The objective is to find a policy in a class of allowed policies $\pi^* \in \Pi$ to maximize the return:
\begin{equation}
\pi^* = \arg \max_{\pi \in \Pi} V_\pi 
\label{eq:main}
\end{equation}

RL algorithms are empirical solutions to MDPs when analytic models for $p$ and $R$ are not precisely known.
In the purest form, the agent starts off with no prior knowledge about how to control the system.
The agent takes random decisions (exploration, and collects a time-series of observations of the states visited, the actions taken, and the effects observed:
\[
x_i = (s,a, r, s')
\]
Different algorithms utilize these observations in differing ways, but the essence is to build a predictive model that finds actions that result in the longest long term benefit (even if the instantaneous reward is small).

\subsection{What is Missing?}
We believe that RL is an important missing piece in OM systems and can facilitate intelligent creation and eviction policy. Materialization is not like paging: there is a complex interplay between immediate effects (use) and long-term effects (the opportunity cost of storing a view).
Even discounting other uncertainty in the DBMS, like errors in query optimizer cost estimation, these are effects that are fundamentally hard to encode as fixed heuristics.

Thus, we advocate for an RL approach that is grounded in real run-times.
We are certainly not the first to consider ``learning'' (or more broadly statistical estimation) in the query optimizer~\cite{markl2003leo, chaudhuri2008pay, kipf2018learned, ortiz2018learning, marcus2018deep, krishnan2018learning}; however, we believe that the problem setting described in this paper is novel.
The first step towards a practical OM system is to develop a framework that learns such a policy through experimentation.

Unfortunately, a direct application of RL will not work. The natural reward function would be the time-improvement for a future query after creating a view. Barring the use of a cost model (Section \ref{exp:cost-model} explains why cost models can be very inaccurate), we cannot directly observe this quantity. Our query optimizer will select to either use or not use a view, thus, there is an unknown ``control'' experiment to accurately evaluate the benefit of creating the view.
Our key algorithmic insight is that RL algorithms can be effectively trained with a series of paired \emph{counter-factual experiments}; how more efficient is the system with a particular view materialized? These experiments can be queued up an asynchronously run.
Of course, this requires an assumption that the full runtime state of the database is easy to reason about. A query run with a different buffer pool state can have a very different runtime.
So, we assume that no such issues exists (e.g., in an in-memory database).
The results of the experiments determine a reward signal that can be fed into an reinforcement learning algorithm.

\section{System Architecture}
In this section we overview our decision model and our system architecture that is integrated with SparkSQL.

\subsection{Decision Model}
Every database $D$ is a collection of base relations (tables) and derived relations (materialized views). 
Let $Q = [q_0,q_1...]$ be an infinite sequence of read-only queries.
These queries are issued to $D$ in the order that they arrive.
As views are materialized and deleted, the database state $D_i$ changes.
The system automatically chooses to re-write queries given the views that are materialized.
Therefore, every query has a latency associated with it, given the current state of the database, and the overall runtime of the workload is defined as:
\[
\textsf{Runtime} =  \sum_{i=0}^\infty \textsf{Latency}(q_i)
\]

For some queries, a suitable rewriting plan will not exist.
So, associated with each query $q_i$ is a set of new views $V_i$ that can be persisted opportunistically (as the query is executed) for the future. To simplify query planning and selection, \emph{we select at most a single view to be persisted at every decision point}.
When we choose to persist a view $v \in V_i$, the overall latency of the query $q_i$ logically decomposes into two parts: 
\[
\textsf{Latency}(q_i) =  \textsf{Cost}(v) + \textsf{Query}(q_i,v),
\]
namely, the cost to create the view $\textsf{Cost}(v)$ and the incremental cost of answering the query with the view $v$. We assume that $\textsf{Query}(q_i,v)$ is stateless, i.e., running the query against different database states does not affect the runtime if exactly the same views are used. This means there are no caching effects or buffer pool effects.

There is a storage cost to persisting such views, which is a function of the current database state:
\[
\textsf{Storage} =  \sum_{i=0}^\infty C(D_i),
\]
and there is further a cap on the amount of storage used at any given time:
\[
 C(D_i) \le \textsf{Capacity}
\]
Therefore, our system possibly needs to evict a view:
\[
D_{i+1} \not \leftarrow v
\]
At each time-step, the state of the database increments based on the view creation and eviction actions taken by the policy $\pi$:
\[
D_{i+1} = \pi(D_i, q_i)
\]

\begin{problem}[Opportunistic Materialization]
Given a database instance $D_0$ and a stream of queries $Q$, plan a set of view creation and eviction operations to:
\[
\min_{\pi \in \Pi} ~~ \sum_{i=0}^\infty \textsf{Latency}(q_i)
\]
\[
\text{subject to: } D_{i+1} = \pi(D_i, q_i)
\]
\[
C(D_i) \le \textsf{Capacity}
\]
\end{problem}

\subsection{Architecture}
\begin{figure}
\centering
\includegraphics[width=0.8\columnwidth]{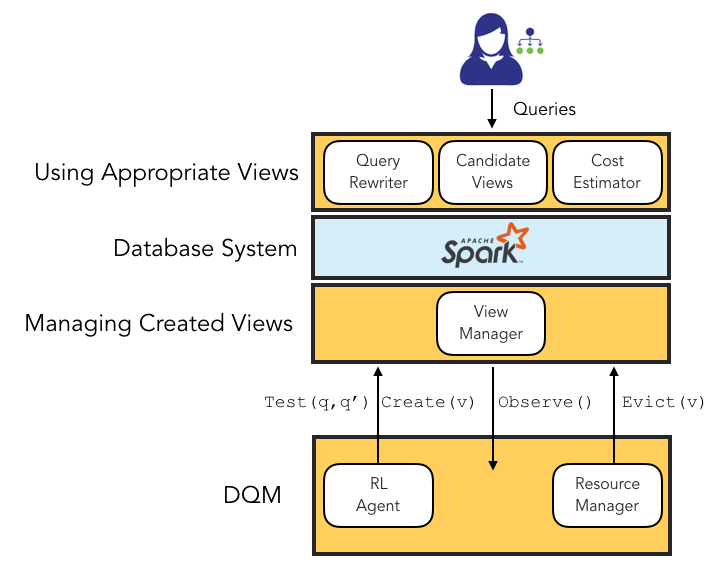}
\caption{\sys runs as an independent process that issues view creation and deletion actions to Apache SparkSQL. A thin wrapper layer around SparkSQL manages the created views and returns any runtime results to \sys. \sys learns from these observations and issues creation and deletion events when appropriate. It also issues potential experiments to run.  \label{arch} }
\end{figure}

We implemented \sys in SparkSQL and overview the architecture in Figure~\ref{arch}. We found that it was convenient to run the OM system in a separate process outside of Spark that issues the creation and deletion actions. All model training occurs asynchronously with another process and is Python. The system interacts with the Spark environment through a RESTful API.

\vspace{0.5em} \noindent \textbf{Current Scope: } \sys currently focuses on inner join predicate views, where every view in the system can be expressed in the following form:
\begin{lstlisting}
SELECT *
FROM R1,...,RN
WHERE C1 AND .... AND CM
\end{lstlisting}
This is not a fundamental limitation of the Deep RL based approach in \sys, but allows for simpler query rewriting to use the views and simpler featurization for views (each view is simply featurized by the scope of its conditions).
We plan to explore more complex views in future work.

\vspace{0.5em} \noindent \textbf{Query Rewriter: }  Given a set of materialized views $V$ and a query $q_i$, the query rewriter changes the query to use the view. A view $v \in V$ is eligible if its predicate conditions are contained in the query. The query rewriter can re-write a query so that any eligible view can be used. Our system searches through all eligible views and selects the lowest cost plan (possibly not using a view at all).

\rulebox{action:materialize-nothing}{}{Set of views $V$, a query $q$}{Rewritten query that uses the view $q_v$}

\begin{figure}[t]
    \centering
    \includegraphics[scale=0.5]{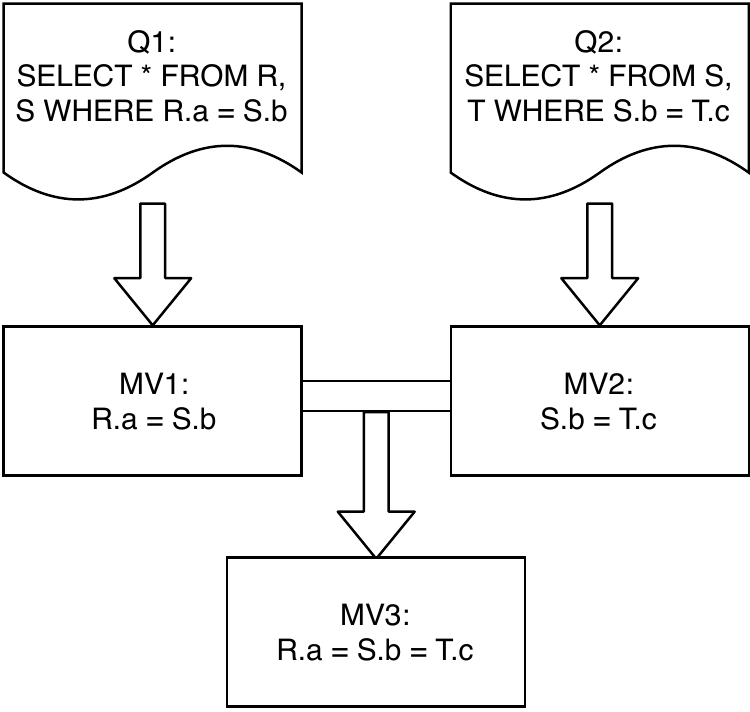}
    \caption{\sys considers opportunistic materialization view candidates whose predicates have been seen at least once before in the workload. We calculate the transitive closure over the equality predicates.}
    \label{mining}
\end{figure}

\vspace{0.5em} \noindent \textbf{View Candidate Miner: } Suppose a query $q$ does not find any useful, eligible views. It can materialize one of its own intermediate results for future use. While in principle, the system could search over all possible views that could be opportunistically materialized while executing any plan of the query, the search space would be prohibitively large. 
As a heuristic, we only select candidate views whose predicates have been seen before in the workload. We calculate the transitive closure over the equality predicates (Figure \ref{mining}).

\rulebox{action:materialize-nothing}{}{A query sequence $Q$}{A set of candidate views $V$}

\vspace{0.5em} \noindent \textbf{View Creation Policy: } We need the system to decide which of the candidate view(s) to materialize. The View Creator decides if and when to create a view from the candidate views:
\rulebox{action:materialize-nothing}{}{A set of candidate views $V$}{A creation action $D \leftarrow v$}

\vspace{0.5em} \noindent \textbf{View Eviction Policy: } Whenever there's room to materialize another view, it makes sense to take full advantage of the available storage space. But once we reach the storage constraint, we need to decide which view to evict to make room for the new view. The View Evictor deletes an already created view in the database:

\rulebox{action:materialize-nothing}{}{The current database $D$}{A deletion action $D \not \leftarrow v$}

\section{Learning Materialization}
In this section, we discuss the core technical contribution of the paper: a reinforcement learning approach for  adaptive view creation.

\subsection{The One View Problem}
Reinforcement learning has received intense research interest in recent years~\cite{sutton2018reinforcement, sutton1992reinforcement}.
We start with a simplified problem and discuss the challenges in making this practical.
Consider the decision of whether to materialize a single view $v$ (ignoring opportunistic selection). Let us  assume there is an automated black-box system process that garbage collects old views when they fall into disuse.
Therefore, our database has a one-bit state--$v$ is currently materialized or not. Our system must simply decide when to apply the unary action to create a view if it's not currently materialized.

While seemingly simple, this decision must actually weigh the benefit of materializing the view in terms of potential query runtime improvement v.s. the creation cost over a workload $Q$:
\[
R(v) = (\sum_{q \in Q} \textsf{Improvement}(q,v) ) - \textsf{Cost}(v)
\]
Views that are not frequently used are not valuable to create.

\subsubsection{Opportunistic Setting}
One might wonder why $\textsf{Cost}(v)$ is relevant to consider in the opportunistic setting, where views are created as artifacts of query execution. Recall from the previous section, that the latency of a query during one of these materialization events decomposes into two terms, the view cost and the incremental query cost using that view:
\[
\textsf{Latency}(q_i) =  \textsf{Cost}(v) + \textsf{Query}(q_i,v).
\]
By explicitly decomposing the problem in this way, we can account for scenarios where the creation of a view may force an instantaneously suboptimal query plan, but creates a view that benefits the cohort of other queries in the long rung.

In our system, ``time-steps'' are synchronized with the query workload, since we only observe an effect when the database is queried. So each query $q_i$ defines a discrete-time decision point of whether to create the view.
However, to be able to apply RL, we need a per-timestep (per-query) reward that quantifies the instantaneous benefit or harm of an action at a particular state.
$R(v)$ is not a well-posed reward function because the creation cost $\textsf{Cost}$ amortizes over the entire workload and does not readily decompose into a per-query value.
We can apply the following trick to produce a consistent reward function:
\[
R(q,v) = \textsf{Improvement}(q,v) - \textsf{Cost}(v) \cdot \frac{\delta(v,q)}{N_v},
\]
where $\delta(v,q)$ is an indicator function determining whether $q$ uses the view or not, and $N_q$ is the number of times the view was used in the past. This means that each relevant query incurs a fractional creation cost. It can easily be verified that\footnote{In practice, we approximately compute the amortization factor $N$, rather than revising rewards retrospectively. Additionally, we can scale $\textsf{Cost}(v)$ by a hyperparameter to adjust unit differences.}:
\[
R(v) = \sum_{q \in Q} R(q,v)
\]

The per-query cost function allows us to model the decision process as an MDP:
\mdpbox{$M = \{0,1\}$ view status, $Q$ workload until $q$ }{$\{\emptyset,+\}$ create the view or do nothing}{$R(q,v)$ improvement minus amortized creation}{$\pi(Q, M) \mapsto \{\emptyset,+\}$ decision to create view}

Our objective is to find a view creation policy: given the current system state (i.e., whether the view is materialized or not and the query workload until the current point), decide the right time to create the view. RL is a framework that learns this policy through trial and error (explore random creation strategies) to optimize the cumulative reward, or $R(v)$ in our case. 

\subsubsection{Counterfacutal Runtime Experiments}
All RL algorithms today assume instantaneous, oracular access to a reward function. 
This is not true in our setting. Evaluating the $\textsf{Improvement()}$ function requires running a query that the system would not ordinarily run, namely, the counterfactual query that doesn't use the view.
This experiment can use a non-trivial amount of system resources.
A key challenge will be to hide this overhead.

A counterfactual scenario is one that is contrary to what actually happened.  Suppose, we make the decision to materialize a view $v$. Then, suppose that $v$ is used to answer a future query $q$ (with observed runtime $\textsf{Query}(q,v)$). There is a hypothetical (counterfactual) world in which $v$ was not created and $q$ was answered without using $v$ (with a counterfactual runtime of $\textsf{Query}(q_i,\emptyset)$).
We are really interested in $\textsf{Query}(q,v) - \textsf{Query}(q_i,\emptyset)$, which is the marginal improvement caused by an action we took in the system:
\[
\textsf{Improvement(q,v)} = \textsf{Query}(q,v) - \textsf{Query}(q_i,\emptyset).
\]

However, we cannot run both queries (with and without the view) in real time as it would expose additional latency to the user.
Our system maintains a running buffer of paired experiments to run.
In idle times, it executes these experiments and stores the marginal improvement for each query.

\begin{figure}
\centering
\includegraphics[width=0.8\columnwidth]{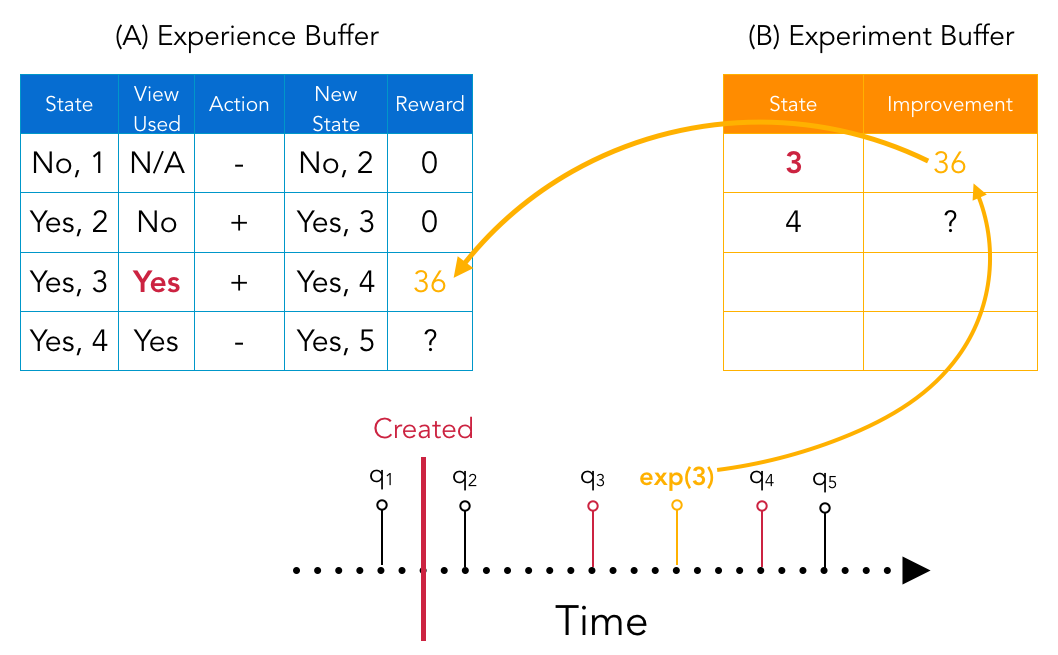}
\caption{We diagram the ``rollout'', or data collection, process used in \sys. At each time-step, \sys decides whether to create a view or not. A reward is received if a created view is used \emph{AND} it improves a query runtime. Every time a created view is used, it queues up an experiment in (B) to run when the system is idle. Once the experiment is run, that observation of improvement is placed in (A) and can be used to improve the policy. \label{data} }
\end{figure}

\subsubsection{Asynchronous RL Algorithm}
The typical anatomy of an RL algorithm is to start with a randomly initialized policy and take decisions to affect the system. It observes the outcomes of its decisions. It periodically retrains the model based on these outcomes making the policy increasingly informed. We now highlight the different parts of our algorithmic framework.

\vspace{0.5em} \noindent \textbf{Rolling out (Data Collection): } The core component of an RL algorithm is the ``rollout'' procedure. Given a policy $\pi$ (whether random or informed), \sys needs to evaluate its effects by applying it to the system. These observations need to be of the form:
\[
(\text{state}, \text{action}, \text{reward}, \text{new state})
\]
As mentioned in the previous section, \sys collects this data in an asynchronous way by maintaining two buffers: an experience buffer and an experiment buffer. Figure \ref{data} diagrams this process.  At each time-step, \sys decides whether to create a view or not. A reward is received if a created view is used \emph{AND} it improves a query runtime. Every time a created view is used, it queues up a counterfactual experiment in (B) to run when the system is idle; run the query with and without the view. 
Once the experiment is run, the observed improvement is placed in (A) and can be used to improve the policy. In short, the experience buffer maintains a set of complete observations.

There is some additional book-keeping that is worth mentioning. Because we only gather such an experience when a view is hit (when a reward is assigned), even though the goal of the algorithm is to select the best view to create, the experience we collected at the time of view hitting is not directly related to view creation. Therefore, there's a gap between a view get created and used, e.g. we create a view at time step T but it only get used at time step T+10. To mitigate the gap between the two events, we tweaked the experience of  $(s, a, r, ns)$ to  $(s', a, r, ns')$ where $s' = s - a$  and $ns' = s' + a = s$. The new experience represents a situation where a view is created and immediately get hit which means the action of creating a view is assigned a reward immediately thus mitigate the gap between view creation and view hitting.

\vspace{0.5em} \noindent \textbf{Policy Update: } Our system continuously collects data and periodically updates the policy. Our RL algorithm is based on the Deep Q Neural Networks (DQN); which as an off-policy algorithm, is robust to asynchronous data collection.
The DQN algorithm defines a \emph{Q-function} (similar to the cost-to-go function):
\begin{equation}
Q(s,a) = R(s,a) + \max_{a'} Q( S',a')
\label{eq:q}
\end{equation}
Given the current state and action, what is the value of this action assume future optimal behavior.
Of course, this function is hypothetical since having this function would imply having an optimal policy.
DQN iteratively approximates this function from data. Let $Q_\theta$ be a parametrized function (e.g., represented by a neural network):
\[
Q_\theta(f_s,f_a) \approx Q(s,a)
\]
where $f_s$ is a \emph{feature vector} representing the state and $f_a$ is a feature vector representing the creation decision. $\theta$ is the model parameters that represent this function and is randomly initialized at the start.
For each training tuple $i$ in the experience buffer, one can calculate the following label, or the ``estimated'' Q-value:
\[
y_i = R_i + \min_{a'} Q_\theta(s',a')
\]
The $\{y_i\}$ can then be used as labels in a regression problem. If $Q$ were the true Q-function, then the following recurrence would hold:
\[
Q(s,a) = R_i + \min_{a'} Q_\theta(s',a')
\]
So, the learning process, or \emph{Q-learning}, defines a loss  at each iteration:
\[
L(Q) = \sum_{i} \|y_i - Q_\theta(s,a)\|_2^2
\]
Then parameters of the Q-function can be optimized with gradient descent until convergence. 

The description above outlines the main theory behind Q-Learning. We also applied the tricks commonly used in practice like Experience Replay and Double DQNs \cite{NIPS2010_3964}.
Experience replay stabilizes DQN training by maintaining a buffer of past observations (rather than truly learning online).
Data are sampled from the buffer for each model update.
The other optimization technique that we use to improve our RL algorithm is called Double DQNs. This technique is used to handle noisy estimates of Q-values. As shown in Equation \ref{eq:q}, we approximate the Q-function by combining the immediate reward and the discounted maximum long term value determined by the DQN itself, which means we are constantly using the DQN to find the best action to take while updating it. The problem is that the best action given by a DQN during updating can be noisy thus could complicate the learning process. To address this problem, the Double DQNs technique suggests using two parametrized neural networks: one network for evaluation the other for updating. We then synchronize the two networks every time an updating process (e.g. 10 epochs) finishes. As a consequence, the learning process become more stable.

\vspace{0.5em} \noindent \textbf{Featurization: } DQN requires that each state and action tuple is featurized. In the 1-view problem, featurization is trivial. It is simply a 1-bit binary vector indicating whether the view is created or not and another bit representing the action to create the view or do nothing.

\begin{table*}[!htbp]
\centering
\begin{tabular}{p{1cm}p{1.5cm}p{3cm}p{1cm}p{2.5cm}p{5.2cm}}
\toprule[0.7pt]
\multicolumn{3}{c}{{\textbf{View Featurization}}} &
\multicolumn{3}{c}{{\textbf{State Action Featurization}}}\\
{View} & {Tables} &{Encoding} & {Action} & {State} &{Encoding}\\
\cmidrule(lr){1-3}\cmidrule(l){4-6}
MV1 & A, B & [1, 1, 0, 0, 0, 0, 0] & MV1 & MV2, MV3 & [1, 1, 0, 0, 0, 0, 0, 1, 1, 1, 1, 1, 0, 0] \\
MV2 & B, C & [0, 1, 1, 0, 0, 0, 0] & MV2 & MV1 & [0, 1, 1, 0, 0, 0, 0, 1, 1, 0, 0, 0, 0, 0] \\
MV3 & A, D, E & [1, 0, 0, 1, 1, 0, 0] & MV3 & MV2, MV4 & [1, 0, 0, 1, 1, 0, 0, 0, 1, 1, 1, 1, 0, 0] \\
MV4 & C, D, E & [0, 0, 1, 1, 1, 0, 0] & MV4 & MV1, MV2, MV3 & [0, 0, 1, 1, 1, 0, 0, 1, 1, 1, 1, 1, 0, 0]\\
\bottomrule[0.7pt]
\end{tabular}
\caption{Featurization of a workload of 7 relations A,B,C,D,E,F,G}\label{one-hot}
\end{table*}

\subsection{Generalizing to N Views}
For simplicity, we introduced the algorithm with a single view to create.
The multiple view case when there is a pool of possible views to create is not that much harder. In principle, we can think of it as N-independent versions of the above algorithm. 

However, there are a few major caveats. First, the set of views may not be known in advance, and could even be dynamic as the workload evolves.  
Second, views might be highly correlated with each other or even mutually exclusive (there is no point creating two very similar views that exclude each other).
So the key change in the N-view version of the RL problem is to additionally record the context of the views, namely, what relations and predicates they consider.

 We take a featurization approach that is similar to \cite{krishnan2018learning}, i.e. we focus on the relations (tables) that are involved in each view, and encode them with one-hot encoding. 
Featurizing an action is straightforward because one action is simply one view. For a state contains multiple views, we perform one-hot encoding on the union of relations involved in all alive views. Finally, we concatenate the action vector and state vector. Table \ref{one-hot} demonstrates our featurization process on a workload that contains 7 relations.

\section{View Eviction}
 In a pure RL setting, it is difficult to enforce a hard constraint, such as a storage limit, with a learned model.
 Therefore, we have to decouple the creation policy from the eviction policy, which independently enforces this constraint.

\subsection{Submissive Eviction}
Our eviction policy ``submissive'' to the RL algorithm in the sense that it's objective is to allow the RL algorithm to act as optimally as possible while enforcing the storage constraint.
Whenever there's room to materialize a desired view, it allows the RL algorithm to make the decision. But if a certain decision exceeds the allotted space, it attempts to free up space such that the constraint can be enforced.

Due to this inherent dependence on the cost of a view and its observed benefit, 
we believe it cannot be addressed by conventional eviction policies: (1) recency metrics like LRU do not account for how beneficial a view is, and (2) most other heuristics do not account for the potential cost of re-materializing a view. In other words, if we already paid the price to materialize an expensive view, then we should evict a cheaper view even if both views brought the same benefit. 
We need an algorithm that is sort of the inverse of the previous RL algorithm; that maintains an estimate of the negative effects of evicting a view.

\begin{figure}
\centering
\includegraphics[width=0.7\columnwidth]{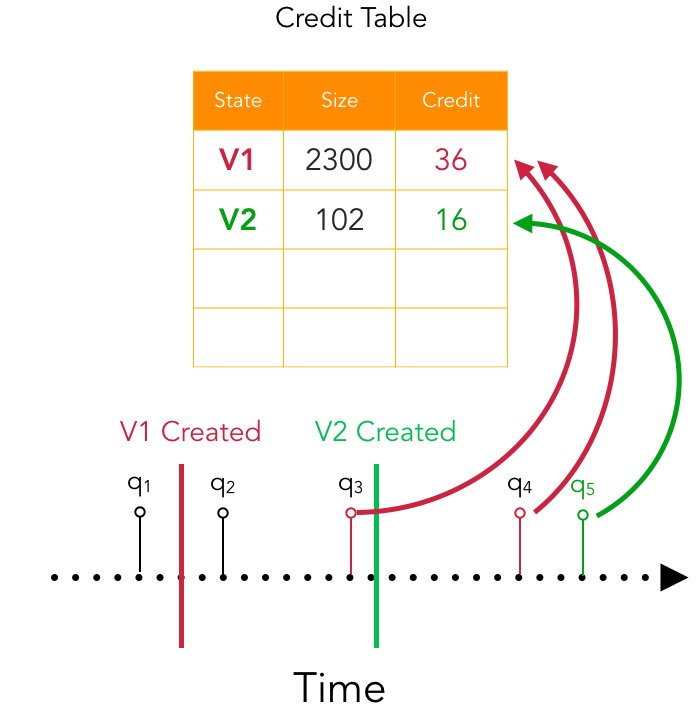}
\caption{\sys maintains a table to prioritize which views to delete to enforce the storage constraint. This table is continuously updated with rewards accrued in the experience buffer. \label{evict} }
\end{figure}

\subsection{Algorithm}
For each created view, the observed value of keeping it materialized is:
\[
R^{-1}(v) =  \sum_{q \in Q} \textsf{Improvement}(q,v) + \textsf{Cost(v)},
\]
or the cumulative improvement so far plus the cost of re-materializing the view. Unlike during creation, where $\textsf{Cost}$ amortizes over each query, there is no such amortization.
If we delete the view there is always a fixed cost of re-creating it.
We maintain a running estimate of its current value as a member of the table (Figure \ref{evict}), each time a view is used by a query:
$$C_{T+1}(v) = C_{T}(v) + R^{-1}(v).$$
Again, as with the view creation policy, we may have to tune hyper-parameters that scale the sum to account for differing units or differing preferences $\textsf{Improvement}(q,v) + \gamma*\textsf{Cost(v)}$.

One challenge is modeling dynamic workloads. If a view was very valuable in the early stage of a workload but then falls into disuse the credit table might have an inflated score.
In practice, we decay the credits of each view by a rate of $\mu \in (0,1]$\footnote{In practice, it is possible for a view to cause negative improvement, we do not decay a negative credit and the hyper-parameter we use to scale the cost will also be negative so that the cost became a penalty instead of a reward to its credit.}:
$$C_{T+1}(v) = \mu \cdot C_{T}(v) + R^{-1}(v).$$
Given the credit table, our eviction policy is to simply evict the view of lowest credit until sufficient space is freed up for the new view.

\subsection{View Maintenance Through Eviction}
We consider an OLAP setting where the materialized views are maintained infrequently. In this problem setting, it is sufficient to treat view maintenance as an automatic eviction event.
For every view currently materialized, if one of its base tables have been updated, we evict it from the pool.
After eviction, we additionally have to flush the experiment buffer of any queued up experiments that use the view since the paired experimental results are now stale.
Since we explicitly model $\textsf{Cost}(v)$ and how it amortizes, our reward function is consistent under maintenance events, as creating a view that is repeatedly evicted will force the view to incur high creation costs that do not amortize well.
This model is sufficient to capture maintenance through re-computation and not incremental view maintenance.
We hope to explore modeling incremental view maintenance in further detail in future work.




\section{Experiments}
\label{expsection}
We explore the following questions: (1) how does \sys compare to conventional heuristics as well as recent state-of-the-art approaches, (2) how quickly does \sys learn a creation effective policy, and (3) how do different hyper-parameters settings affect \sys.

\subsection{Setup}
We implemented \sys in SparkSQL. Our RL algorithm is decoupled from the Spark environment and is implemented using the Keras framework\cite{chollet2015keras}. All experiments are run on a cluster of machines each with 2 Intel E5-2680 2.40 GHz CPUs and 64G memory running Scientific Linux 7.2. We run each experiment 5 times and present the average of the 5 runs.

\subsubsection{Workloads}
\label{sec:workloads}
Queries and data are derived from the Join Order Benchmark (JOB) and TPC-DS. JOB is based on the IMDB dataset and consists of 113 aggregate queries with joins of up-to 16 tables. TPC-DS is based on a synthetic dataset and a query generator that generates queries with aggregates, joins, and subqueries. Our TPC-DS data is generated with a scale factor of 1.

We generate different scenarios from these two benchmarks. Our default workload is called \textsf{para}.
\textsf{para} is a steady state workload (does not evolve with time) that does not contain a skew of frequency on specific queries.
Queries from both benchmarks are augmented with random single-attribute predicates, so that the exact same query never appears twice. 
The workloads contain a sequence of 1000 such queries and the queries are submitted and served in a sequential manner. 
Queries arrive at regular intervals, and the asynchronous experiments can be run in a single time-step (we evaluate these effects explicitly later).

Then in subsequent experiments, we go beyond \textsf{para} and apply the power laws skew the frequency of queries as previous work has done~\cite{drapeau1994toward, lo2010generating}. \textsf{dzipf} skews both query generators to run more expensive queries with a much higher frequency, and \textsf{azipf} skews the query generators to run the least expensive queries with a much higher frequency.
However, for both these workloads the query frequency is fixed throughout the 1000 queries, albeit skewed.
Next, we consider dynamic workloads.
 \textsf{dablend} is a 1000 query workload that starts of executing the most expensive queries then switches to executing the least expensive queries, and \textsf{adblend} does the opposite.

We normalize the storage constraint across all experiments. The available storage for opportunistic materialization is set to 200MB (roughly 20\% the size of the largest base table in the experiments). This allows us to compare results across workloads in an apples-to-apples way. 

\begin{table}[t]
\small
\centering
\begin{tabular}{p{1.5cm}p{6.2cm}}
\toprule[0.7pt]
\textbf{Name} & \textbf{Description}\\
\midrule
azipf & Rank the queries by latency in ascending order then apply the zipf distribution.\\
dzipf & Rank the queries by latency in descending order then apply zipf distribution.\\
rzipf & Shuffle the queries then apply zipf distribution.\\
adblend & Concatenate the first 500 queries from the $azipf$ workload with the first 500 queries from the $dzipf$ workload.\\
dablend & Concatenate the first 500 queries from the $dzipf$ workload with the first 500 queries from the $azipf$ workload.\\
para & Parameterize the queries with unique random parameters.\\
\bottomrule[0.7pt]
\end{tabular}
\caption{Macro-workloads characteristics}\label{workloads}
\end{table}

\subsubsection{Baselines}
We implemented a number of baselines shown in Table~\ref{baselines} that range from conventional cache algorithm, like Least Recently Used (LRU), to sophisticated heuristic-based approaches from previous work.
All of the baselines benefit from the other components of \sys such as the candidate view miner.
\sys proposes relevant views that can be opportunistically generated by the current query and relevant to the past workload.
The baselines have to select which of these views to persist and evict existing views if necessary.

\begin{table}[t]
\small
\centering
\begin{tabular}{p{2.0cm}p{5.7cm}}
\toprule[0.7pt]
\multicolumn{2}{c}{\textbf{Baselines}}\\
\hline \\
\hline
\multicolumn{2}{c}{Eviction Only}\\
\hline
LRU & Randomly select one of the candidates, evict the least recently used view in the cache.\\
LFU & Randomly select one of the candidates, evict the least frequently used view in the cache.\\
FIFO & Randomly select one of the candidates, evict the earliest view inserted into the cache.\\
\hline 
\multicolumn{2}{c}{Selection and Eviction}\\
\hline
HAWC~\cite{perez2014history} &  Select the best view from the candidates based on the Spark query optimizer cost model. For each materialized view, maintain a ``credit table'' based on subsequent query cost that uses the view (cost difference of using vs. not using the view). The credit table is windowed to take the latest $K$ queries. HAWC evicts the lowest credit view. \\
RECYCLER~\cite{recycler6544837} & Select the most expensive view (in terms of creation cost). Materialize a new view if its cost is higher than existing views. Evict the lowest cost view otherwise. For views in the cache, the cost is scaled up when used, scaled down when not used. Our default implementation of Recycler makes use of the true costs of views, we further study a more practical alternative using cost model estimated view costs in Section \ref{value_of_truth}. \\
\hline 
\multicolumn{2}{c}{Hypothetical}\\
\hline
BELADY$^*$ & Select the optimal view to use based on complete, accurate foresight and use Belady's algorithm\cite{belady1969} to evict old views.\\
\bottomrule[0.7pt]
\end{tabular}
\caption{Baselines used in the paper}
\label{baselines}
\end{table}

\begin{figure}[t]
    \centering
    \includegraphics[scale=0.2]{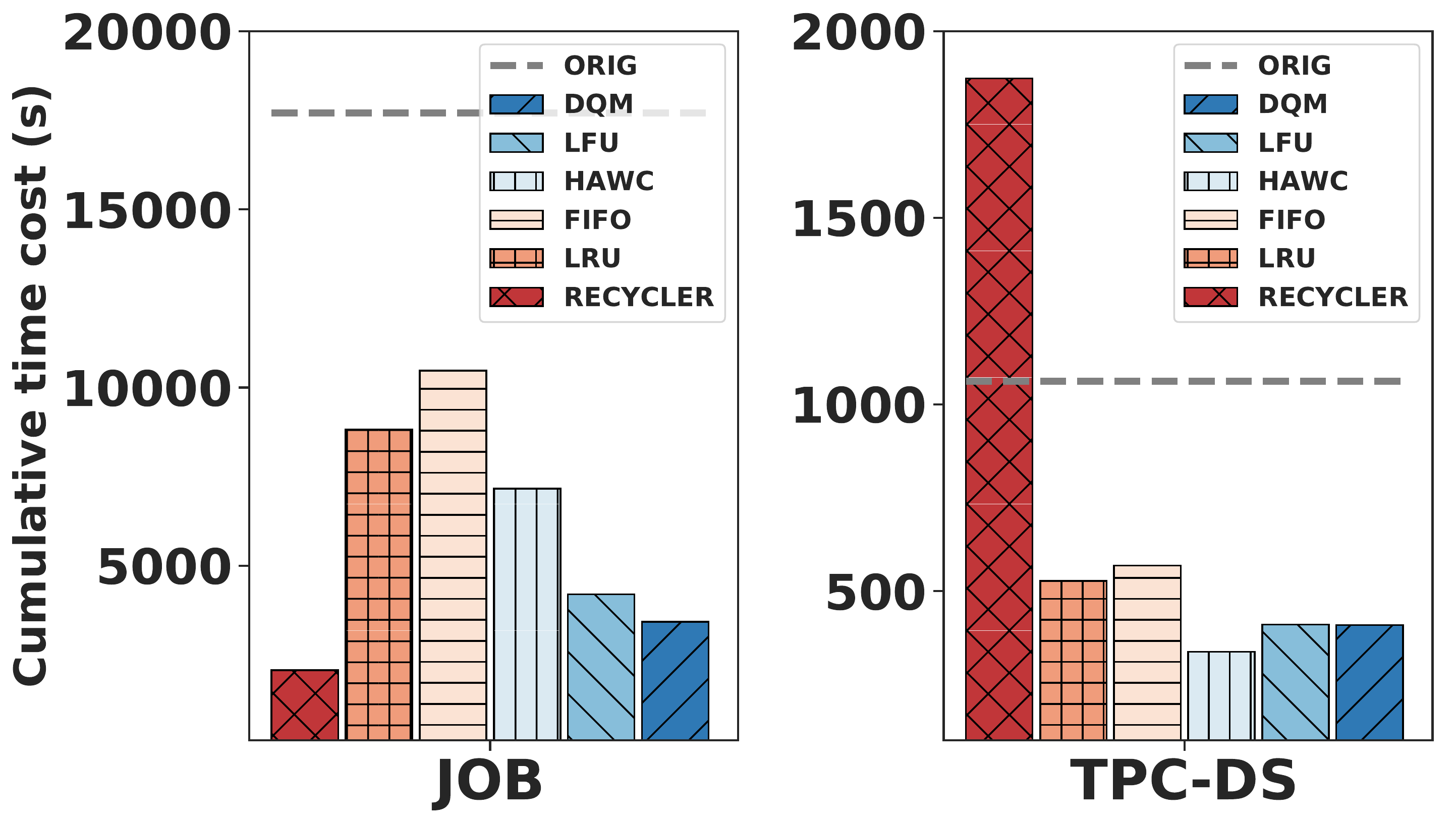}
    \caption{We compare \sys to all of the baselines on the \textsf{para} workload. Even including learning time, \sys is competitive with the best baselines in both TPC-DS and JOB.}
    \label{para-all}
\end{figure}

\subsection{Baseline Performance}
We first evaluate \sys and the baselines on \textsf{para} (Figure \ref{para-all}).
We measure the cumulative runtime of the entire 1000 query workload. The neural network of \sys is initialized randomly and has to learn the creation and deletion policy online.
\emph{This exploration time for \sys is included in the overall runtime.}

As described in the introduction, Recycler works well when its creation cost heuristic correlates with improvements in runtime.
Recycler speeds up query latency in JOB by over 10x. 
There is a significant amount of nuance in these results. 
We provide Recycler with an \emph{exact} cardinality estimate for the size of the views.
While this is possible to know in hindsight after the views are created in order to prioritize deletions; it is impossible to know this exactly during creation time (i.e., a join cardinality estimation problem). Nonetheless, we are generous to Recycler as future experiments show that a faulty cardinality estimate very significantly affects results.

The next interesting insight from this experiment is that all of the ``eviction-only'' strategies perform reasonably well on both benchmarks. Randomized selection with a sensible eviction heuristic leads to up-to a 3x improvement on both benchmarks. 
 HAWC uses an informed selection policy based on Spark's query optimizer but its drastic performance shift on the two workloads indicates optimizer based selection policy is not reliable, we further discuss how such an inexpensive estimation would affect \sys in Section \ref{value_of_truth}.

\sys is competitive with all baselines on both benchmarks even when it has to learn. By leveraging real runtime observations, it is robust to cost estimation issues in the query optimizer. To us, this is a very surprising insight. There is overhead in the exploration process as the system has to learn from suboptimal actions. Even so, the system is competitive with the best baselines during this learning phase. We will also see that \textsf{para} is a worst-case of sorts for \sys.

\subsection{Skew Performance}
\begin{figure*}[ht]
    \centering
    \includegraphics[width=0.48\linewidth]{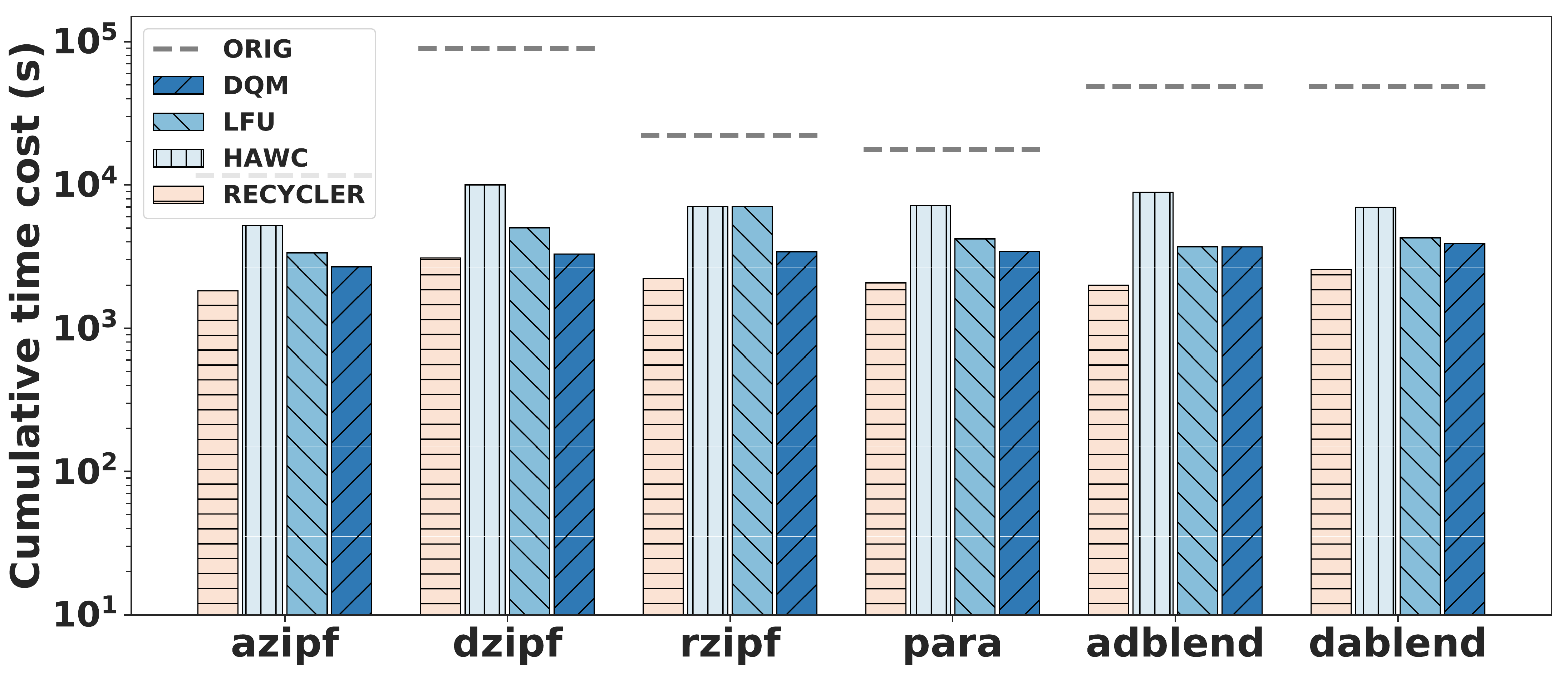}
        \includegraphics[width=0.48\linewidth]{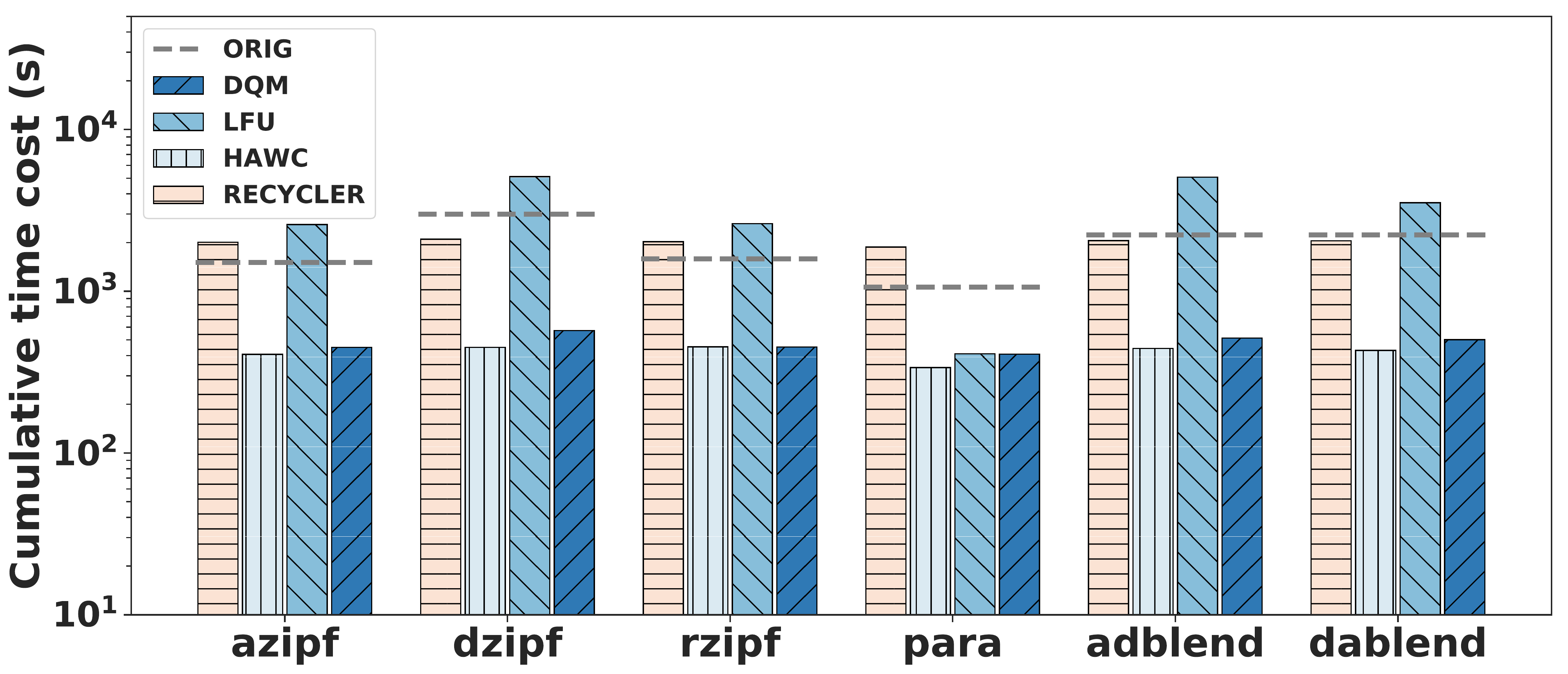}
    \caption{We compare \sys to selected baselines on all workloads. \sys is competitive (or outperforms) the best heuristic on all the test scenarios.}
    \label{allw-J}
\end{figure*}

We dig deeper on these baselines and consider different query skews and query distributions. We evaluate \sys, LFU and the two heuristic-based approaches with all 12 different workloads. Results are shown in Figure \ref{allw-J} A and Figure \ref{allw-J} B for JOB-based workloads and TPCDS-based workloads respectively.
We find that the results from the previous experiment broadly hold across all of the different skews.

Recycler works very well on JOB-based workloads, it outperforms \sys on JOB-based workloads. On all of these workloads, Recycler stops admitting new views after processing the first 100 queries. However, this heuristic has significant issues on the TPC-DS workloads. When we dug into the TPC-DS results, we found the view miner generated a single view candidate repeatedly that was very large, and confusing enough to the query optimizer that it hurt the performance when used by some queries. 


We find that HAWC's performance also changes drastically on JOB and TPCDS. This is because its view selection is based on the inaccurate estimation of Spark's optimizer which we also demonstrate in Section \ref{value_of_truth}.

\subsection{Maintenance Performance}
\label{maintenance_section}
The heuristics break down when there are costs that they do not model or anticipate. 
Maintenance costs in OLAP systems are infrequent but are significant.
In this experiment, we study how view maintenance could affect \sys and the baselines. Because Spark does not support incremental update of views, every time the base tables are modified we have to re-compute and re-materialize the views that are affected. To simulate periodic maintenance, our system will randomly select a base table of the workload and evict all views using the table. 

For comparison purpose, we perform a eviction at every 100 queries, and this is controlled so all approaches have the same maintenance routine. Even though the maintenance routine is the same, a different approach will introduce different maintenance cost because different views are materialized. 

\begin{figure}[ht]
    \centering
    \includegraphics[scale=0.18]{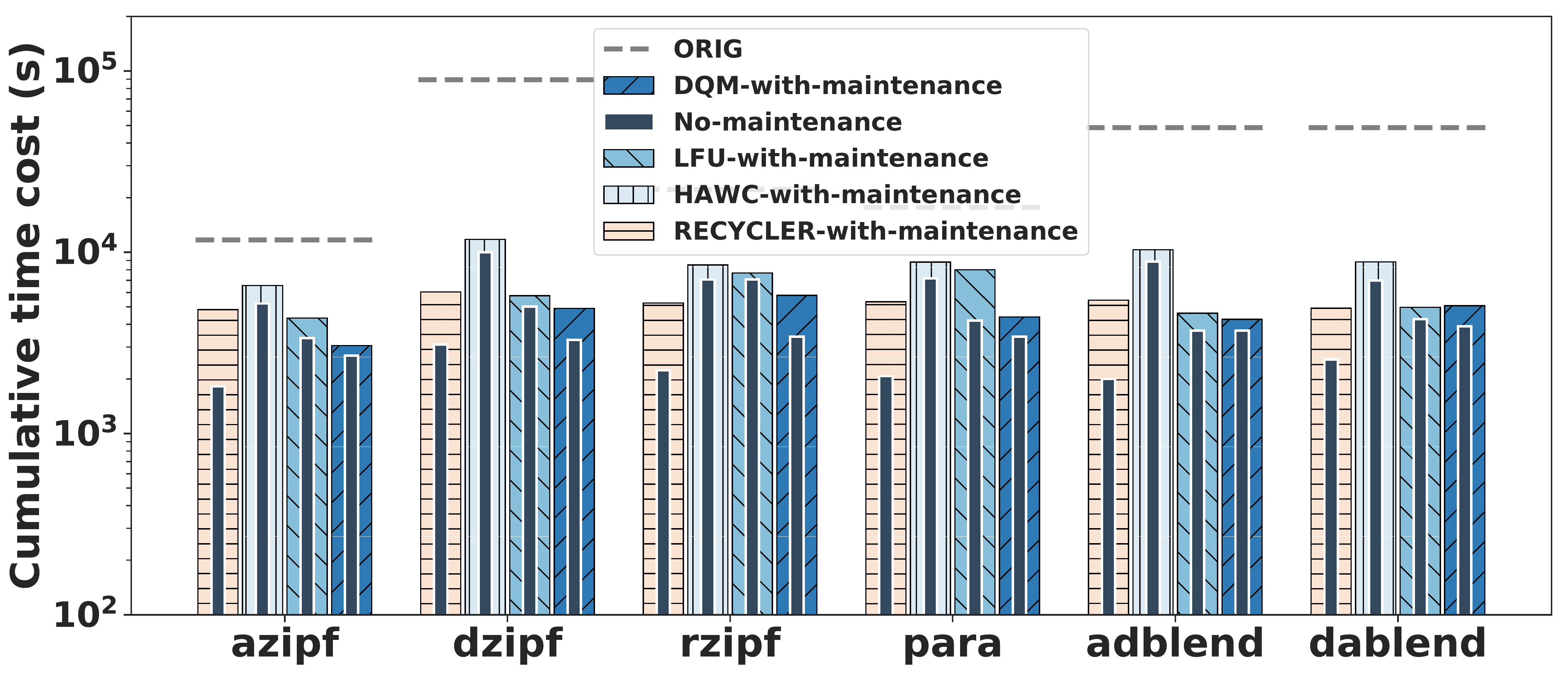}
    \caption{We compare \sys to selected baselines with periodic view maintenance (every 100 queries) on the JOB workloads.}
    \label{maintenance}
\end{figure}

Results can be found in Figure \ref{maintenance} for the \textsf{para} workload on the JOB benchmark. Maintenance certainly adds an overhead to all techniques, but the results demonstrate that \sys is more efficient and robust to maintenance.
In the previous experiment, we found that Recycler was very effective on this workload. But after maintenance, we found \sys now outperforms Recycler non-trivially on 4 of the 6 JOB-based workloads because \sys.  Recycler's heuristic favors expensive views thus selecting views that incur a higher maintenance overhead than \sys. Again, the benefit of \sys is a direct optimization of observed query latencies. Views that have to be constantly recreated because they are maintained fall out of favor of the learning algorithm quickly.

\subsection{Exploration vs. Exploitation}
We evaluate \sys in a pure online setting. At the beginning of each workload, \sys has to deal with the cold-start problem. \sys starts with random selection to explore until the number of experience and as observations come it, it periodically re-trains its model.
As we collect more observations, we become more confident about \sys, and then we start to explore less with random actions. The exploration parameter is $\epsilon$; $\epsilon$ represents the exploration rate and $1-\epsilon$ represents the probability of exploiting what we have learned. 

On the other hand, we must always have some degree of random creation to ensure that \sys is adaptive to changes. Our system starts from an $\epsilon = 1$ (always take random actions) and decays this value to $\epsilon\_min$.
 We evaluate \sys using different $\epsilon\_min$.

\begin{figure}[ht]
    \centering
    \includegraphics[scale=0.2]{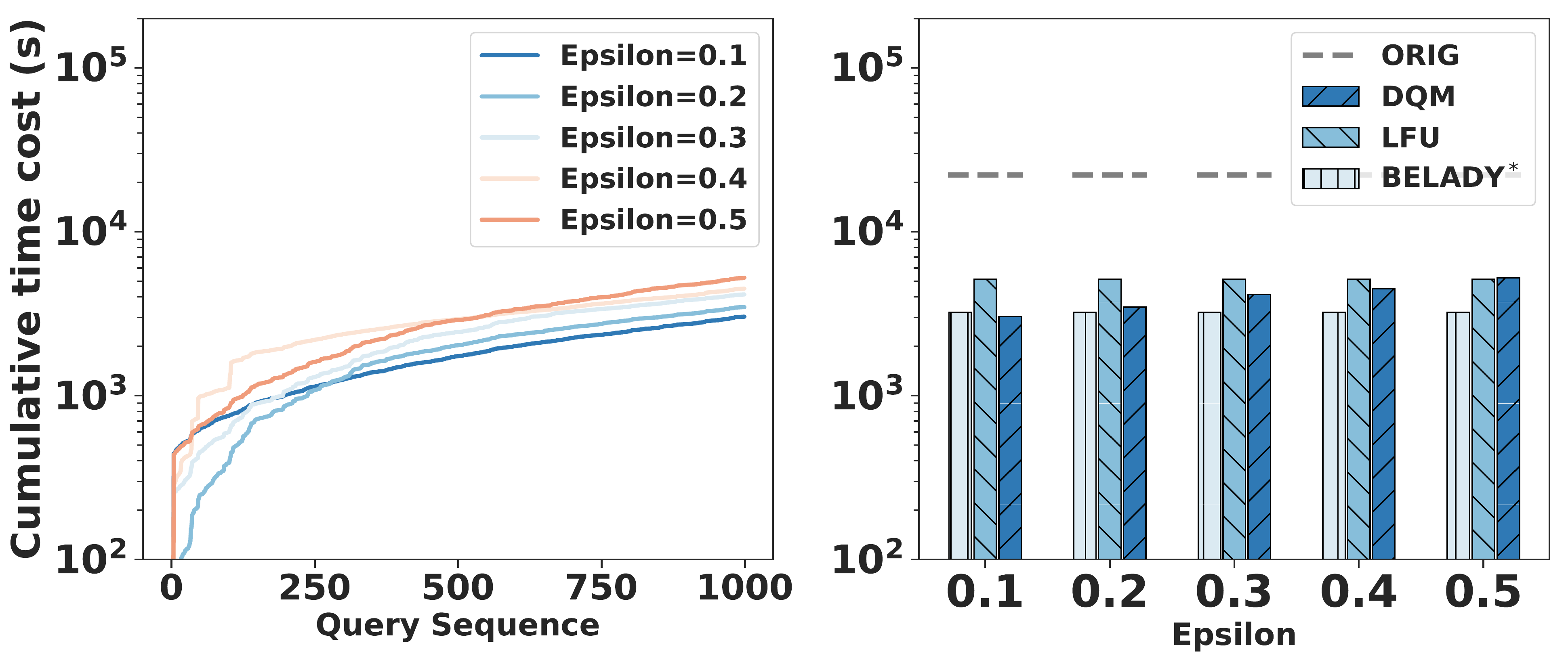}
    \caption{\sys using different exploration terms on JOB with the rzipf workload.}
    \label{epsilon}
\end{figure}

We set $\epsilon\_min$ to 0.1, 0.2, 0.3, 0.4, 0.5 and results are shown in Figure \ref{epsilon}\footnote{Other experiments use a fixed $\epsilon\_min$ of 0.1.}. As expected, even though exploration is necessary for \sys to avoid settling on a local optimal, as we learn more we should prefer more exploitation and a high $\epsilon\_min$ hurts the performance. However, even with a high $\epsilon\_min$ of 0.5, \sys still performs competitively with LFU. Curiously, a higher exploration term benefits \sys in the early stages of the workload (such as 0.2). 

\subsection{Cost Estimation Issues}\label{exp:cost-model}
\label{value_of_truth}
In our previous results, we were actually very generous to our baselines.
We provide the baselines with exact view cardinality estimates (\sys does not use this as it learns purely from observed runtimes). 

\begin{figure}[ht]
    \centering
    \includegraphics[scale=0.22]{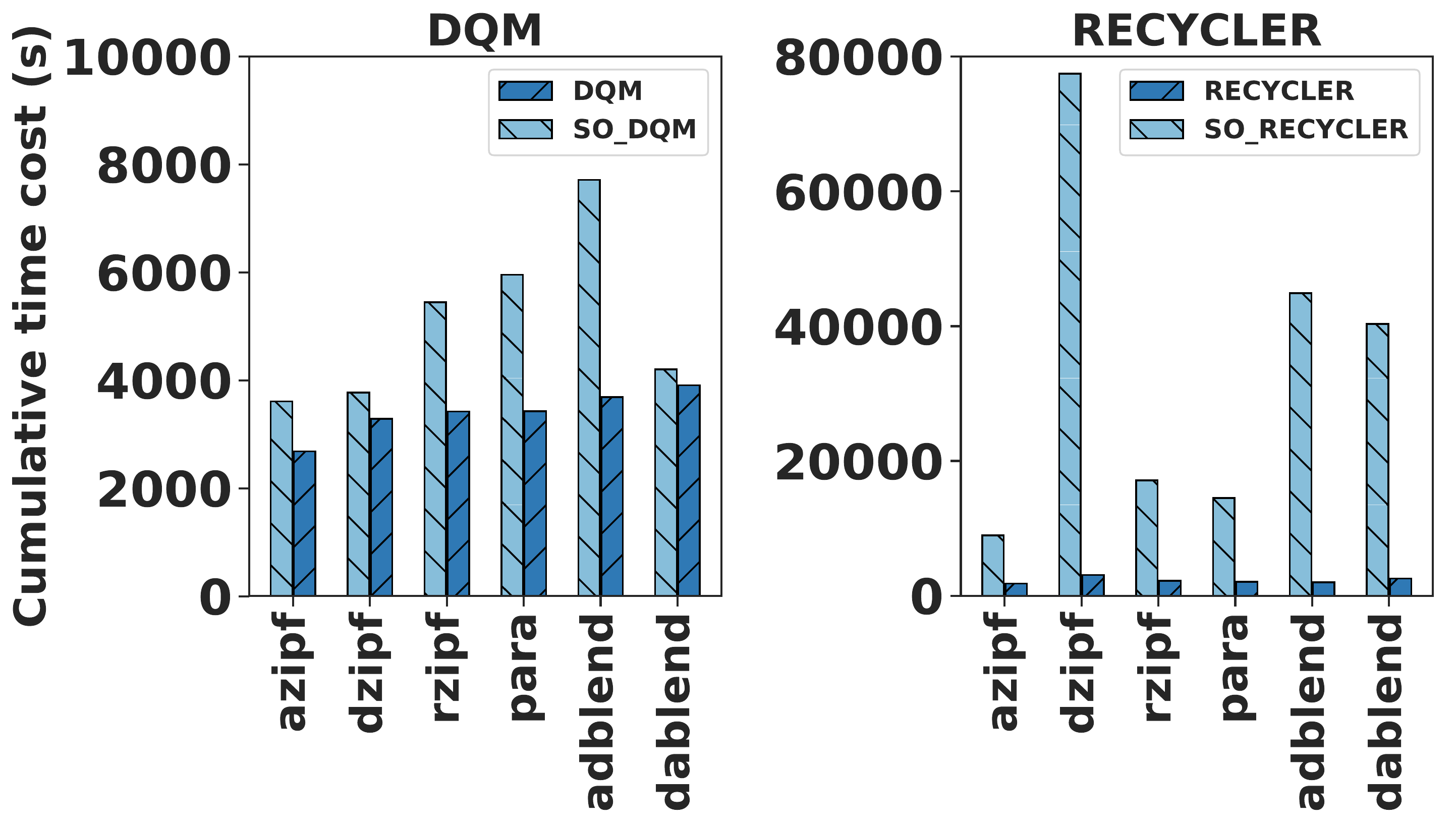}
    \caption{We run \sys without true runtimes and an improvement metric derived from a cost model. While this version of \sys still performs reasonably well, the use of true runtimes is a strength of the RL-based algorithm. We also evaluate a more realistic implementation of Recycler by using the cost model estimated view costs instead of the true costs of views and this change drastically affects Recycler's performance.}
    \label{sodqm}
\end{figure}

The Recycler baseline is most sensitive to this. We implemented a more realistic alternative of Recycler called SO$\_$Recycler. The only difference between the two is that SO$\_$Recycler uses SparkSQL's query optimizer to estimate the cost of a view instead of using the true cost. As we can see in Figure \ref{sodqm}, this change significantly affects Recycler's performance because the costs of views play an important rule in Recycler's heuristic: it assumes that a more expensive view will bring more benefits. Therefore, when the costs of views are inaccurate its performance drops up to 25x on the \textsf{dzipf} workload.

We could do the opposite with \sys; what is the effect if we use Spark's optimizer for an inexpensive cost estimate rather than the counter-factual experiments. The difference in query cost using or not using the view can be used to determine the improvement.
In this experiment, we modify \sys to use estimated reward from SparkSQL's query optimizer to investigate how it would affect the performance of \sys. We call the Spark optimizer based version $SO\_$\sys and results can be found in Figure \ref{sodqm}.
As mentioned in earlier sections, reward functions play an important role in RL systems, it is designed to guide the model towards the direction of the highest long term value. Therefore, it is not surprising an RL system underperforms when its reward function is inaccurate or even wrong. In practice, this gap is up to a factor of 2x in the \textsf{adblend} workload. 
$SO\_$\sys is still reasonable in its performance but we believe that the power of RL is to feedback true execution times. Directly optimizing the true reward function explains much of the power of \sys.

\subsection{Storage Constraints}
\begin{figure}[ht]
    \centering
    \includegraphics[scale=0.18]{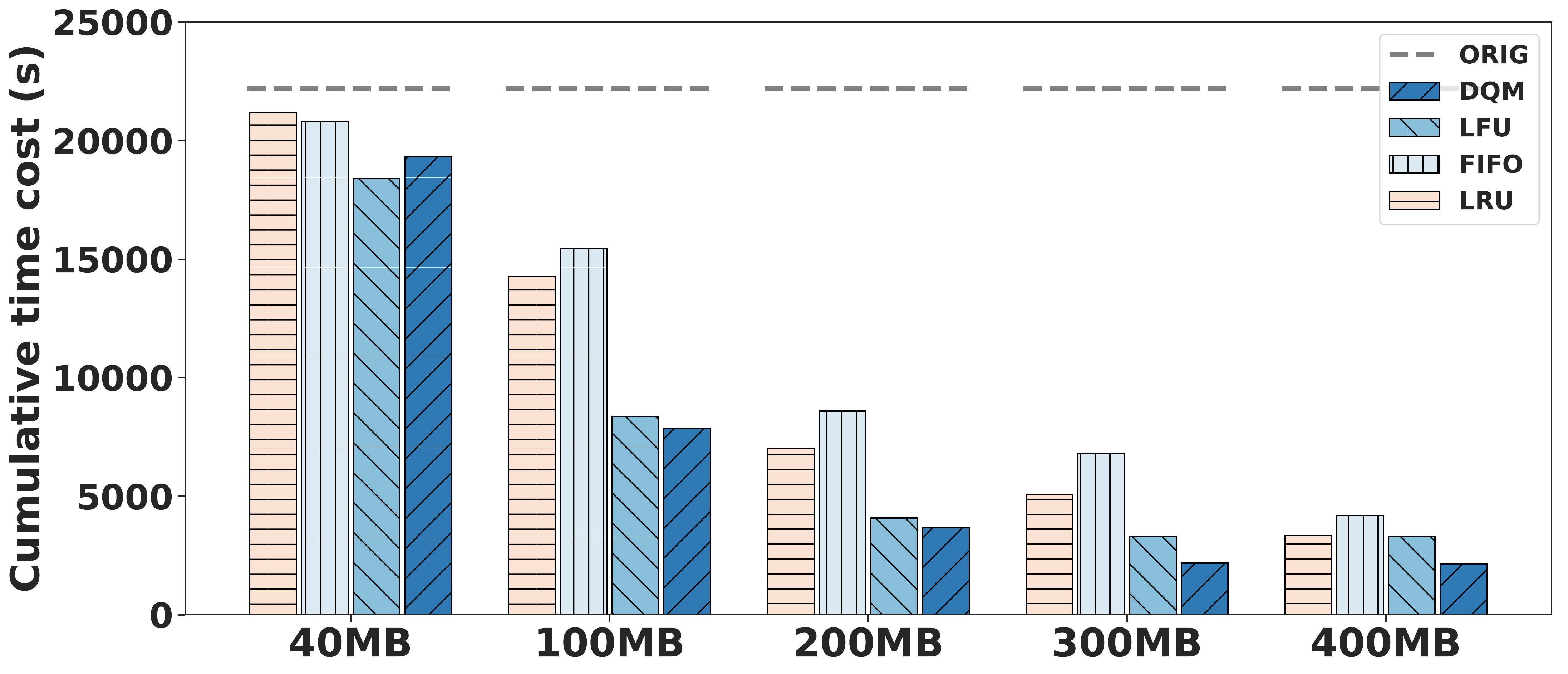}
    \caption{We measure the performance of \sys as a function of the storage constraint on the JOB rzipf workload. The storage constraint is presented as normalized by a fraction of candidate views that could possibly be materialized at any time.}
    \label{cachesize}
\end{figure}

To study how \sys reacts to changes in the storage constraint, we use the same JOB-rzipf workload from the previous experiments. The result is shown in Figure \ref{cachesize}, where the storage constraint is normalized by a fraction of candidate views that could possibly be materialized.

OM is most valuable when there is a substantial amount of spare storage in the system. The power of OM is trading off this spare storage for future query latency. As expected, the performance of \sys and baselines improve as we increase the storage constraint. We note that the most significant increase is between 40MB and 100MB\footnote{For reference the 200MB datapoint is the level used for \sys and all baselines in the previous experiments.}. 

\subsection{Delayed Rewards}
 In this experiment, we explore how delays in the asynchronous experimentation affect \sys. \sys relies on system idle time to execute paired experiments, what if this idle time is contended?
 We simulate this in the following way: given a delay of $K$, an experience that is generated at time step $T$ will only be available to \sys for learning at time step $T+K$.
 As an extreme example if $K$ equals 1000, \sys will select views completely randomly for our 1000 query workloads.
 
This is a worst case simulation of delayed reward.
In practice, the system idle time will likely be more randomly distributed and some queries might get earlier observations. However, by pushing all the experiences to the end of the process, we are simulating the worst case of delayed rewards. I.e. if the same amount of experiences were thrown away but system idle time was more randomly distributed, then more experiences will be available earlier for the agent to learn from thus benefiting the system.

\begin{figure}[ht]
    \centering
    \includegraphics[scale=0.18]{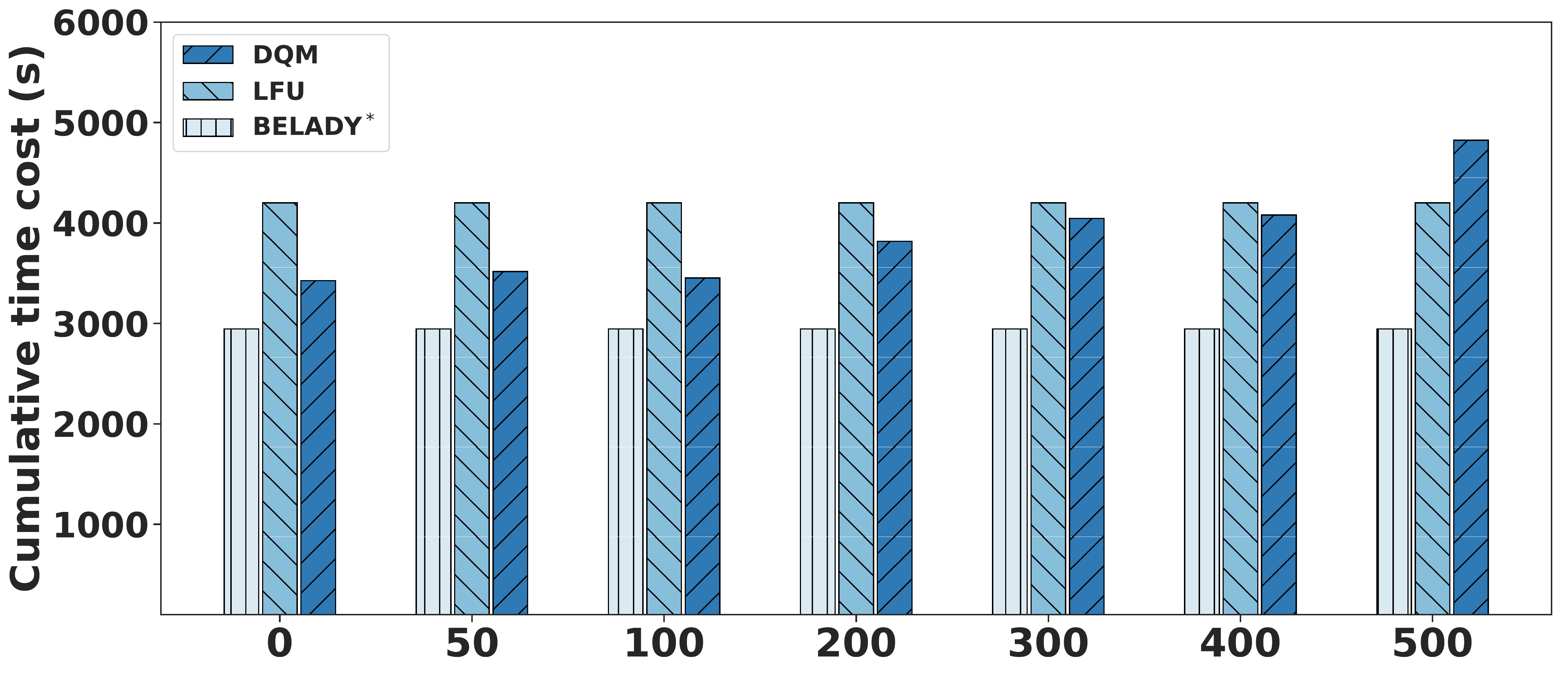}
    \caption{We measure the performance of \sys using different delays of reward on JOB para workload.}
    \label{delay}
\end{figure}

We use the \textsf{para} workload from JOB in this experiment and test delay of 50, 100, 200, 300, 400 and 500.
The results can be found in Figure \ref{delay}. We see that the performance of \sys changes drastically with a delay of 500 because half of the experiences are thrown away. A delay under 200 does not affect much of \sys's performance, it still outperforms LFU non-trivially. 

\subsection{DQM vs. Belady$^*$}
Next, we want to understand how well \sys is doing in absolute terms.
Belady$^*$ is a hypothetical baseline whose eviction policy is based on the Belady's algorithm\cite{belady1969}, which is hypothetical because it relies on hindsight to evict a view that will not be needed for the longest time in the future. Belady$^*$ is also hypothetical because it always selects the best view (from an oracle) in foresight. Therefore, Belady$^*$'s admission policy is optimal, but its eviction policy is not optimal because it does not consider the cost and benefit of a view. We believe that it would be prohibitively expensive to optimize over all possible optimal creation AND deletion policies. 

\begin{figure}[ht]
    \centering
    \includegraphics[scale=0.22]{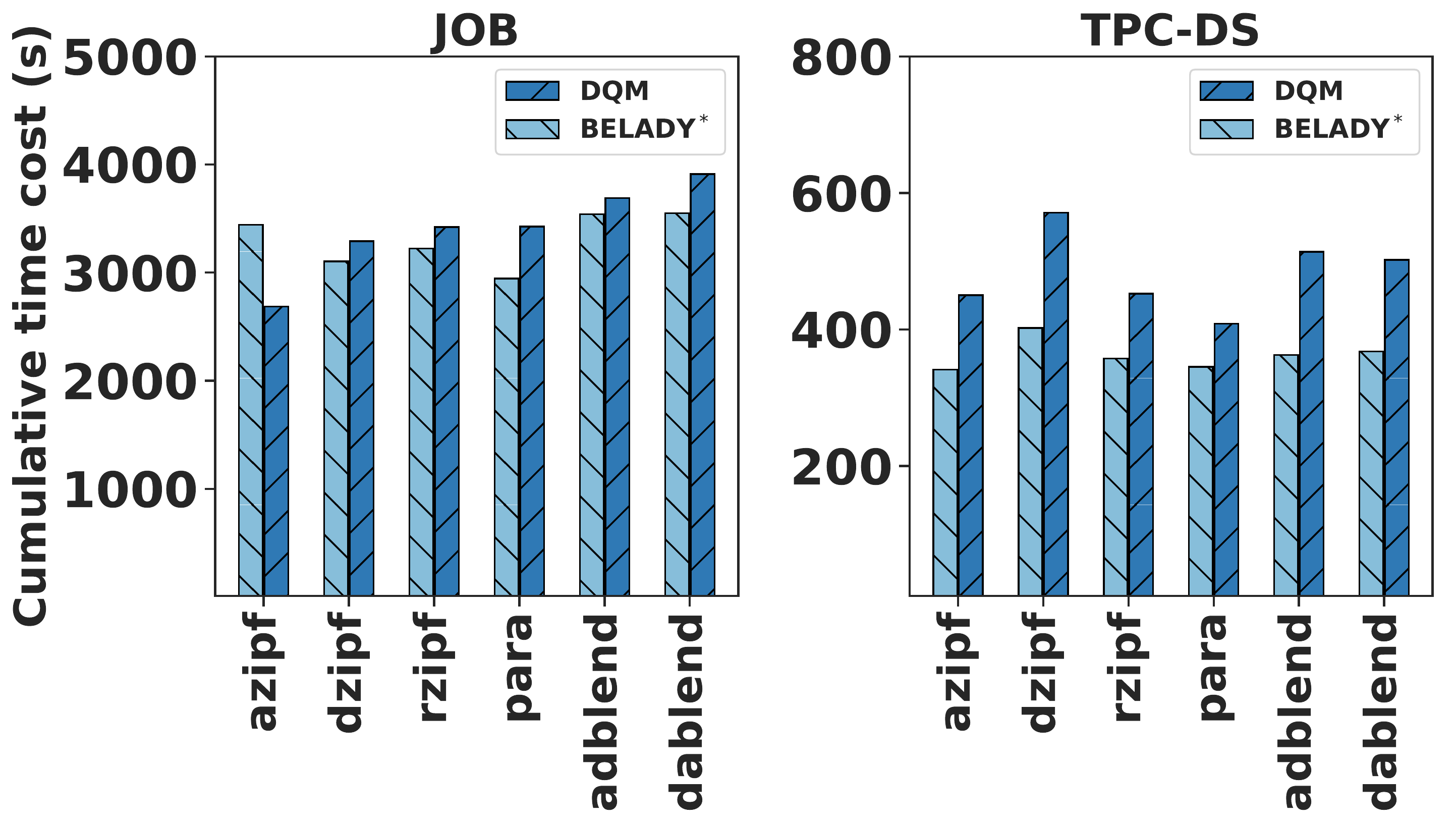}
    \caption{We compare \sys to a hypothetical near-optimal solution called Belady$^*$, which uses perfect foresight to select the best views and evict a view that will not be needed for the longest time in the future (Belady's algorithm).}
    \label{lfuopt}
\end{figure}

We evaluate Belady$^*$ on all 12 workloads. By learning a predictive model that directly optimizes for runtime improvement, \sys actually performs competitively with Belady$^*$ on both Job-based workloads and TPC-DS workloads as shown in Figure \ref{lfuopt}. \sys learns a stateful creation policy that considers what views are already materialized and a cost-aware eviction policy informed by those improvement experiments. In the rest of this section, we evaluate \sys on a set of micro-benchmarks to understand how different settings affect \sys's performance.

\subsection{Overhead}
In all of our previous experiments, we \emph{included} the overhead of learning and exploration in our cumulative runtimes.
Next, we try to understand the system performance after learning (assuming that the future workload is stationary, of course).
 We first train \sys with  1000 queries from the \textsf{para} workload using a $\epsilon\_min$ of 0.1 then we change $\epsilon\_min$ to 0 (to avoid exploration), and then, use the trained model to select views for another 1000 queries (with no additional exploration).

\begin{figure}[ht]
    \centering
    \includegraphics[scale=0.15]{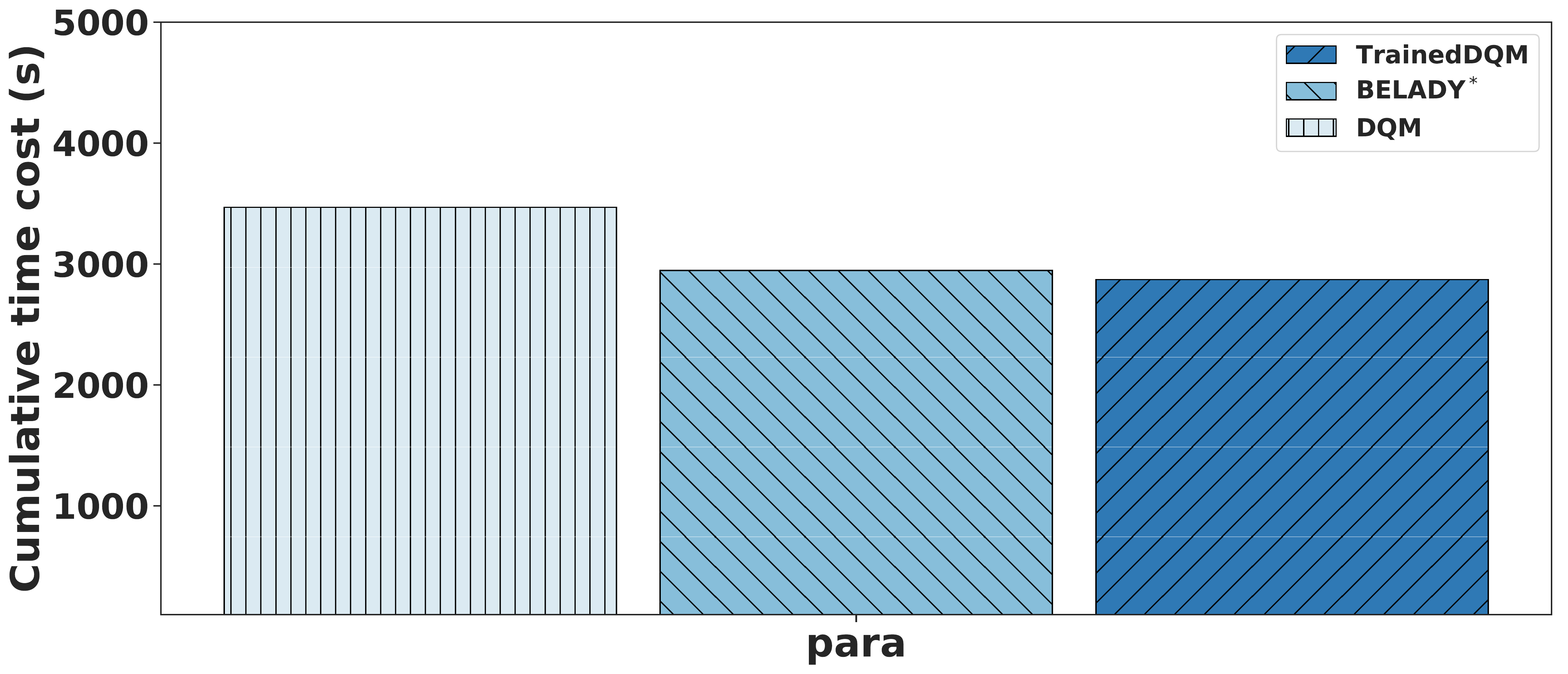}
    \caption{We evaluate overhead caused by learning and exploration by comparing \sys with a Trained\sys with no exploration.}
    \label{learning_overhead}
\end{figure}

The result can be found in Figure \ref{learning_overhead}. The over all performance improves by about 15$\%$. This shows that \sys is a very data-efficient learning approach. The paired experiments give a very strong signal for learning, and \sys is able to quickly learn from this signal.

It also shows that continuously learning is not too onerous for a real system. It is advantageous to always have some amount of exploration (some seemingly suboptimal decisions). This allows the system to ``bounce'' out of local minima if the workload or the execution environment changes.
However, that said, we do believe  that overfitting can be an issue--if this was the case, the difference between \sys and a Trained\sys would be more significant if Trained\sys erroneously memorized patterns of the workload (if any). We leave further investigation of this to future work as we build and deploy \sys in realistic scenarios.

\section{Discussion and Future Work}
We believe one limitation of our current work is a sophisticated methodology for query featurization. A better query featurization will definitely help the agent learn better and make better decisions, e.g. capture latent patterns in the workload. This problem is related to the work studied by Ortiz et al. in the context of query optimization~\cite{ortiz2018learning}.
We also believe that more work can be done studying relatistic dynamic workloads. We are actively looking for benchmark workloads that simulate ad hoc querying.  We believe such workloads are necessary for building and evaluating more practical methodologies.

There are also numerous opportunities for reusing computation and intermediate query state in OLAP workloads, and we believe that machine learning will be an important part of future OLAP systems.
Applications of machine learning in database internals are still the subject of significant debate, and will continue to be a contentious question for years to come~\cite{btree, kraska2018case, ma2018query}. An important question is what problems are amenable to machine learning solutions. We believe that materialization is one such sub-area. 

We see \sys as a first step towards a View-Oriented database, one that aggressively anticipates future queries and materializes anything that could be useful. Such an architecture shifts the query optimization burden from planning a query to efficiently reusing past computation. New algorithms and theory will have to developed to understand the new problem setting. We believe machine learning will be an important part of this discussion.
In the short-term, extending \sys to consider OLTP settings and more complex reward functions is certainly a priority. We also want to explore dynamic or periodic workloads.

\balance


\bibliographystyle{abbrv}
\bibliography{bibs}  








\end{document}